\documentclass[12pt]{report}
\usepackage[english]{babel}
\usepackage[utf8]{inputenc}
\usepackage{amssymb}
\usepackage{hyperref}
\usepackage{setspace}
\usepackage{mathtools}
\usepackage{graphics}
\usepackage{subcaption}
\usepackage{tikz}
\usepackage{xspace}

\usepackage{scrextend}
\usepackage{afterpage}

\usepackage{amsmath}
\usepackage{breqn}

\usepackage{float} 
\usepackage{graphicx}
\usepackage{booktabs}
\usepackage{array}
\usepackage{setspace}

\pdfoutput=1

\PassOptionsToPackage{normalem}{ulem}
\usepackage{ulem}
\providecolor{added}{rgb}{0,0,1}
\providecolor{deleted}{rgb}{1,0,0}

\title{\textbf{A Review on Recommendation Systems: Context-aware to Social-based}}

\author{\textbf{S.M. Mahdi Seyednezhad} \\ Department of Computer Engineering and Sciences\\ Florida Institute of Technology\\ Melbourne, Florida, USA.\\sseyednezhad2013@my.fit.edu\\ \\
\textbf{Kailey Nobuko Cozart} \\ Department of Mathematics and Computer Science\\ Whitworth University\\ Spokane, Washington, USA\\ cozart21@my.whitworth.edu\\ \\
\textbf{John Anthony Bowllan} \\ Department of Mathematics\\ Middlebury College\\ Middlebury, Vermont, USA
\\jbowllan@middlebury.edu\\ \\
\textbf{Anthony O. Smith} \\ Department of Computer Engineering and Sciences\\ Florida Institute of Technology\\ Melbourne, Florida, USA.\\anthonysmith@fit.edu
}

\usepackage{atbegshi}
\AtBeginDocument{\AtBeginShipoutNext{\AtBeginShipoutDiscard}}

\usepackage{xcolor,soul,lipsum}

\newcommand{\myul}[2][black]{\setulcolor{#1}\ul{#2}\setulcolor{black}}

\begin{document}
\begin{titlepage}
\pagestyle{empty} 
\maketitle
\end{titlepage}
\clearpage
\pagenumbering{arabic} 
\tableofcontents

\definecolor{red}{rgb}{1,0,0}
\onehalfspacing

\let\Oldcite\cite
\renewcommand{\cite}[1]{~\Oldcite{#1}}
\newcommand{\et}{et al.\xspace}
\newcommand{\TFIDF}{\ensuremath{\textit{TF-IDF}}}


\chapter*{Abstract}
The number of Internet users has grown rapidly enticing companies and cooperations to make full use of recommendation infrastructures. Consequently, online advertisement companies emerged to aid us in the presence of numerous items and users. Even as a user, you may find yourself drowned in a set of items that you think you might need, but you are not sure if you should try them. Those items could be online services, products, places or even a person for a friendship. Therefore, we need recommender systems that pave the way and help us making good decisions. This paper provides a review on traditional recommendation systems, recommendation system evaluations and metrics, context-aware recommendation systems, and social-based recommendation systems. While it is hard to include all the information in a brief review paper, we try to have an introductory review over the essentials of recommendation systems. More detailed information on each chapter will be found in the corresponding references. For the purpose of explaining the concept in a different way, we provided slides available on \href{https://www.slideshare.net/MahdiSeyednejad/recommender-systems-97094937}{\color{blue} \myul[blue] {Slideshare}}.
\addcontentsline{toc}{chapter}{Abstract}

\chapter{Introduction}
\label{ch:introduction}

It is unbelievable that the human brain has evolved to deal with our complex world. However, this same world has recently been augmented by technology and humans are dependent on such technology to perform daily tasks. This vicious cycle means that humans now require outside help to make sense of the world, in particular input from others regarding their previous experience; that is, their recommendations. 

Surprisingly, during ancient civilizations (4000 - 1200 BC), humans needed recommendations. They could be used for problems such as what crops to cultivate, the appropriate time of cultivation, what religion to follow, etc.
After that, in old times (and probably nowadays) families used to recommend acquaintances to each other for arranged marriages. Currently, people ask for recommendations regarding many aspects of modern life such as travel, music, movies, etc. The idea of recommendation systems began to be more important after the industrial revolution in which the number of available goods grew enormously and it became vital when computers changed the global market\cite{sharma2016evolution}.

By 2015, the number of Internet users had grown from 738 million in 2000 to 3.2 billion\cite{itu:users}, meaning that 43\% of the world population was using online services, enticing companies and cooperations to make full use of recommendation infrastructures. Consequently, online advertisement companies emerged to aid us in the presence of numerous items and users. Even as a user, you may find yourself drowned in a set of items that you think you might need, but you are not sure if you should try them. Those items could be online services, products, places or even a person for a friendship. Therefore, we need recommender systems that pave the way and help us making good decisions.

Recommender systems have attracted the attention of a significant number of popular Internet sites, such as Amazon.com, YouTube, Netflix, Spotify, LinkedIn, Facebook, Tripadvisor and IMDB\cite{linden2003amazon}. Particularly, many media companies offer practical recommender systems to their subscribers. Based on the type of applications, there are various purposes for a recommender system, including but not limited to, increasing the number of items sold, selling more diverse items, and increasing user satisfaction and fidelity\cite{RS_book_ch1_2015}. 

Recommender systems (RSs) collect information from users' preference for a set of items and predict the best desired items for them\cite{bobadillaSurvey2013}. The information can be obtained explicitly by recording users' ratings or implicitly by observing users' behavior. Generating a recommender system depends on a set of considerations, such as type of available data, filtering algorithms, models, techniques, sparsity level of data and desired quality\cite{bobadillaSurvey2013}. Some recommender systems are designed for a specific task. For instance, Guan et al.\cite{Guan2016} introduced a recommender system for apparel.
Recommender systems found their way to becoming an independent research area in the mid 1970s at Duke University\cite{sharma2016evolution}. 

Recommender systems are used with a lot of information about items, users, and ratings. In an information filtering system, unwanted information is removed by using computerized methods prior to presentation to the users. 
Its main goal is the to manage the information overload and to increase the semantic signal-to-noise ratio\cite{hanani2001information}. In fact, a recommender system needs to filter information in order to find more relevant items for the users.
Demographic filtering, content-based filtering, collaborative filtering and hybrid methods are the main four methods of recommender systems \cite{candillier2007comparing,adomavicius2005toward}. Among them, collaborative filtering (CF) and the methods combining with it are the most popular ones because they are based on user ratings\cite{salter2006cinemascreen,herlocker2004evaluating}. Content-based filtering is based on content of the items that the users liked in the past\cite{van2000using}. 
For example, if the user tried a science fiction movie in the past, the recommender system will most likely recommend a recent science fiction movie\cite{pazzani1999framework}. This method is popular for websites such as IMDB, Rotten Tomatoes, and Pandora. On the other hand, in demographic filtering, the recommender system observes the common attributes of the users (gender, age, location) and suggests items to the users with similar attributes. This is based on the principle that people with some specific common attributes may have common interest. 

When we want to take into account user ratings, we should make use of collaborative filtering. These systems try to predict the utility of items for a specific user based on the items that the user previously rated\cite{adomavicius2005toward}. These days, massive online companies such as Amazon, Facebook, Twitter, LinkedIn, Spotify, Google News and Last.fm employ this technique.
The most widely used algorithm for collaborative filtering is k Nearest Neighbors (kNN)\cite{bobadilla2011framework}; its application is based on two main approaches: user to user and item to item. In the user to user version, the kNN algorithm first tries to determine the k neighborhood for the user; then, it aggregates users based on their ratings and finally predicts based on the aggregated information. Gong \cite{gong2010collaborative} uses both items and users to implement a bi-cluster method for a new recommender system. The major pitfalls of using the kNN algorithm for recommender systems are the high level of sparsity in RS datasets and its low scalability\cite{luo2012incremental}.

Cold start is one of the main challenges that almost all recommender systems face when the initial ratings or any knowledge about user experience is not sufficient. There are three major cold-start problems\cite{bobadillaSurvey2013}: new community, new user and new item. In the new community problem, the RS suffers from a lack of sufficient data when it is initialized for a new community. This becomes even harder when the RS uses a pure collaborative filtering based on community preferences\cite{schein2002methods}. In Chapter~\ref{ch:methods}, we provide more details about recommender system filtering methods. In Chapter~\ref{ch:context}, we talk about the concept of context in recommender systems and the major approaches for deploying the contextual information in a recommender system. The emergence of online social networks raises the concept of context. In Chapter~\ref{ch:socialInfo}, we provide additional information on recommender systems when the social information plays an important role.

\chapter{Recommender System Methods and Evaluation}
\label{ch:methods}
As we mentioned in Chapter~\ref{ch:introduction}, there are three traditional main recommendation methods: demographic filtering, content based filtering and collaborative filtering\cite{pazzani1999framework}; among them, demographic and content-based filtering have been the most popular ones. However, most of the big companies primarily use a hybrid approach which is a combination of the aforementioned methods\cite{bobadillaSurvey2013}. In this chapter, we explain these methods in more detail and discuss each one's virtues and drawbacks. 

In general, the information about an active user's feedback is crucial for recommender systems and it is obtained explicitly or implicitly. Explicit feedback is the users' ratings on items, and it can be considered as direct information about feedback. The main advantage of this type of feedback is its simplicity; nonetheless, the drawback is the need for active users to rate items. Unfortunately, some users do not rate items. On the other hand, implicit feedback is extracted by monitoring user behavior and analyzing user activity. For example, if a user tries action movies frequently, the implicit information implies that the user's rating on action movies could be high. In the case that an item is a document, then, printing, saving, reading or bookmarking could be the reflect of user interest in that document. The distinct advantage of this method is that there is no need for an active user to rate items; however, sometimes biasing can happen. For example, if you are interrupted by a phone call while you are opening a document, the recommender system may judge you as a fan of that document, which may not be true.

\section{Demographic Filtering}
This type of recommender systems suggests items based on the demographic profile of users. It can be used to identify the taste of users that belong to a certain community. Therefore, to design these systems, we need some information about users to categorize them into groups. Then, if some users in a particular group like or order an item, it is possible that the other users of this group tend to do the same. It should be noted that although it might be better to use more structured information about users, there is a trade-off between the computational complexity and the quality of demographic filtering. Pazzani\cite{pazzani1999framework} ran an experiment based on demographic filtering on data about restaurants and he claimed that on average, 57.5\% of the top three recommended restaurants were liked by users.
Table~\ref{tab:demographic} shows an example of people who rated a specific restaurant. It tells us that a female in an area in which the code is 714 is probably going to like the restaurant.

\begin{table}[ht]
\caption{Demographic information on the users who rated a specific restaurant.}
\begin{center}
  \resizebox{0.8\textwidth}{!}{%
  	\tabcolsep=0.2cm
    \begin{tabular}{l c c c c c r}
     \toprule
      User & Gender & Age& Area Code & Education & Employed & Rate \\
       \midrule
       Karen & F  & 15 & 714  & HS & F & +  \\
       Lynn & F  & 17 & 714  & HS & F & $-$  \\
       Chris & F  & 27 & 714  & C & T & +  \\
       Mike & M  & 40 & 714  & C & T & $-$  \\
       Jill & F  & 10 & 714  & E & F & ?  \\
      \bottomrule
    \end{tabular}%
    }
\end{center}
\label{tab:demographic}
\end{table}

\section{Content-based Filtering}
\label{sec:content}
Content-based methods make recommendations based on the description of the items. Nowadays, it is combined with other methods and use more information about items and users. However, several algorithms have been proposed to analyze the content of a document. For example, we consider the case in which a recommender system is designed with the content-based method to recommend movies to users. We may assume that the movie description has been already extracted. If the movie is an action film and a user liked it, the recommender system will recommend another action movie to the user. 

Content-based recommender systems (CBRS) consist of three major parts from a high level architectural point of view\cite{marco2015content}; first it does the preprocessing on items with a {\it content analyzer}, then a {\it profile learner} learns about users. Finally, the {\it filtering component} finds a set of appropriate recommendations. More details for these three parts are provided as follows:

\begin{itemize}
\item {\bf Content Analyzer} --- For any decision making problem, the raw data should be pre-processed to extract featured information. Here, the output of this pre-processing part is the structured relevant information. The content analyzer prepares information for the next step. It transforms information from its original format to one that is more abstract and useful. For example, it may receive a web page as input and convert it to a vector of keywords. 

\item {\bf Profile Learner} --- This module is specifically designed for the user side. It receives the pre-processed information from the content analyzer and generalizes them to construct the user preferences. The generalization step models the user interest based on the user's past ratings of items. For example, a profile learner in a web page recommender system may combine the vector of positive and negative examples to construct a {\it prototype} item vector that represents the user profile\cite{salton1997improving}.

\item {\bf Filtering Component} --- This is the final part that finds the relevant items based on the user profile and recommends them to the user. It uses a similarity measure (e.g. cosine similarity) between items and the user prototype.
\end{itemize}

Content-based recommender systems are mainly used where the item is either a document or a text used to describe an item. Thus, text mining methods play an important role in content-based recommender systems. The traditional methods are highly sensitive to the way that documents are represented\cite{pazzani1999framework}. However, the following technique is widely used to analyze the content of a text document and turn it into a vector.

\subsection{Keyword-based vector space model}
The Vector Space Model (VSM) is one of the spatial representations of text documents. It transforms a text document to a $n$-dimensional space (i.e. a vector with $n$ elements) in which each dimension (element) is a term in the given document collection. This method needs to weight the terms and calculate the similarity of documents based on those weights. The most commonly used weighting method for terms is TF-IDF (Term Frequency Inverse Document Frequency) weighting, which is based on information extracted from text.

In TF-IDF, terms that are frequently found in one text (TF), but rarely in other documents (IDF), will be possibly more related to the topic of that text\cite{marco2015content}. Additionally, the weight is normalized to give the same chance of being retrieved to both small and large documents. Equation~\ref{eq:tf} shows TF based on term frequencies.

\begin{equation}
TF(t_{k},d_{j}) = \frac{f_{k,j}}{max \{ f_{z,j} \} }
\label{eq:tf}
\end{equation}

\noindent where $t_{k}$ denotes the $k$th term in the dictionary of terms $T=\{ t_{1},t_{2},t_{3}, ... , t_{n} \}$, $d_{j}$ is a document from the document collection $D=\{ d_{1},d_{2},d_{3}, ... , d_{N} \}$, $f_{k,j}$ is the frequency of term $t_{k}$ in document $d_{j}$, and $max \{ f_{z,j} \} $ is the maximum of all the frequencies of all terms in document $d_{j}$. Moreover, we have equation~\ref{eq:idf} to calculate the IDF based on the size of the collection and the documents with a particular term $t_{k}$.

\begin{equation}
IDF(t_{k}) =\log \frac{N}{n_{k}}
\label{eq:idf}
\end{equation}
\noindent where $N$ is the number of documents (collection size), and $n_{k}$ is the number of documents in which the term $t_{k}$ has been seen at least once.

Equation~\ref{eq:tf-idf} uses the obtained TF and IDF to calculate the TF-IDF for each term in each document.
\begin{equation}
\TFIDF(t_{k},d_{j}) = TF(t_{k},d_{j}) \cdot IDF(t_{k})
\label{eq:tf-idf}
\end{equation}

Now, we need to normalize the weights to be in $[0,1]$ and to have the vectors with the same length. Equation~\ref{eq:wkj} does the cosine normalization for this purpose.

\begin{equation}
w_{k,j} = \frac{\TFIDF(t_{k},d_{j})}
{\sqrt{\sum_{s=1}^{\left | T \right |} { {\TFIDF(t_{k},d_{j})}^{2} }
}}
\label{eq:wkj}
\end{equation}

\noindent where $w_{k,j}$ denotes the weight corresponding to term $t_{k}$ in document $d_{j}$. 

One similarity measure is needed to test the closeness of two documents. Equation\ref{eq:sim} calculates the similarity between documents $d_{i}$ and $d_{j}$ using cosine similarity, which is common in this field.

\begin{equation}
sim(d_{i},d_{j}) = \frac{\sum_{k} {w_{k,i} \cdot w_{k,j}}}
{ \sqrt{\sum_{k} { {w_{k,i}}^2 } } \cdot \sqrt{\sum_{k} { {w_{k,j}}^2 } } 
}
\label{eq:sim}
\end{equation}

In content-based filtering methods that use VSM, both user profiles and items are represented by vectors of weighted terms\cite{marco2015content}. Recently, semantic aware methods have attracted the attention of scholars who have been working on content based recommender systems.
\section{Collaborative Filtering}
We consider a recommender system for movies.The RS may face a situation in which we do not know a particular movie's features, but we know how some specific users rated it. Now, if two users named ``{\it Marcos}'' and ``{\it Diego}'' like a movie titled {\it A}, and later {\it Marcos} watches another movie titled {\it B} and likes it, then we can recommend this movie to {\it Diego}. This approach is adopted from the collaborative filtering method.

In collaborative filtering, the recommender system looks for similarity between users to make predictions. In several cases, the pattern of ratings of users is a useful feature to determine similarity\cite{pazzani1999framework}. Normally, collaborative filtering recommendation methods use patterns of ratings or usage to recommend items specified for users without need for extra information about either users or items\cite{koren2015collab}. Similarly to other recommendation methods, CF methods must relate items and users which are two essential different entities. The
{\it Neighborhood approach} is a technique that concentrates on how items or users are related among themselves. For instance, in an item-item approach, the RS models the preference of a user to an item with respect to the previous rating of the same user to a similar item. Another technique is the {\it latent factor model}, such as matrix factorization, which transforms both items and users to the same latent factor space.

Additionally, there has been a different point of view to categorize the collaborative filtering techniques by dividing it into {\t memory-based} and {\it model-based} methods. Memory-based methods act only on a user-item rating matrix and can easily be adapted to use all the ratings before the filtering process; thus, its results are updated. On the other hand, a model based system, like a neural network, generates a model that learns from the information of user-item ratings and recommends new items\cite{roh2003collaborative_ann}. 

In memory-based CF methods, measuring the similarity plays a significant role, because the RS either tries to find the similarity between items or the similarity between users\cite{Su2009ColabSurvey}. It needs to find the similarity between items to see what a user's opinion is of items and what the closest new/unseen/unknown item is to the items that the user has already liked, following which a recommendation can be made. Likewise, it needs the similarity between users to see what are the close users, and if a user tries a new item, the RS recommends it to the users close to her.

Among various similarity measures, we mention the {\it Pearson Correlation measure}, which reveals the information on how much two variables are linearly related to each other. Equation~\ref{eq:wuv_pearson} calculates the Pearson correlation between user $u$ and $v$ which gives us the information about the similarity of users who both rated the same item.

\begin{equation}
w_{u,v} = \frac{\sum_{i \in I_{u,v}} { (r_{u,i}-\bar{r}_{u}) \cdot (r_{v,i}-\bar{r}_{v}) }}
{ \sqrt{\sum_{i \in I_{u,v}} { {(r_{u,i}-\bar{r}_{u})}^2 } } \cdot \sqrt{\sum_{i \in I_{u,v}} { {(r_{v,i}-\bar{r}_{v})}^2 } } 
}
\label{eq:wuv_pearson}
\end{equation}

\noindent where $I_{u,v}$ is the set of items that both users $u$ and $v$ rated, $\bar{r}_{u}$ is the average rating of items rated by user $u$ in $I_{u,v}$, and $r_{u,i}$ is the rating of user $u$ on item $i$.

In a similar fashion, by using Pearson Correlation in equation~\ref{eq:wuv_pearson}, we can calculate the similarity of items $i$ and $j$ that were rated by users: 

\begin{equation}
w_{i,j} = \frac{\sum_{u \in U_{i,j}} { (r_{u,i}-\bar{r}_{i}) \cdot (r_{u,j}-\bar{r}_{j}) }}
{ \sqrt{\sum_{u \in U_{i,j}} { {(r_{u,i}-\bar{r}_{i})}^2 } } \cdot \sqrt{\sum_{u \in U_{i,j}} { {(r_{u,j}-\bar{r}_{j})}^2 } } 
}
\label{eq:wuv_pearson}
\end{equation}
\noindent where $\bar{r}_{j}$ is the average rating of item $i$, and $U_{i,j}$ is the set of users that rated both items $i$ and $j$.

After computing the item-item and user-user similarity, the RS job is to predict a rating on a particular item from a certain user. In a neighborhood-based model, a nearest neighbor should be picked to be involved in predicting the ratings. If we assume that we have an active user $a$ and that the RS needs to predict the user's rating on item $i$. In {\it weighted sum of others' ratings}, the predicted rating is calculated by equation~\ref{eq:wuv_Pai}.
\begin{equation}
P_{a,i} =\bar{r}_{a}+ \frac{  \sum_{u \in U} {(r_{u,i}-\bar{r}_{u}) \cdot w_{a,u}}    }
{\sum_{u \in U} |w_{a,u}|}
\label{eq:wuv_Pai}
\end{equation}

\noindent where $P_{a,i}$ is the predicted rating of user $a$ for item $i$, $\bar{r}_u$ is the average rating of user $u$, $\bar{r}_a$ is the average rating of item $a$, and $w_{a,u}$ is the weight between those two users calculated by equation~\ref{eq:wuv_pearson}. Additionally, we can define a threshold for $w_{a,u}$ to avoid participating the considerably small weights.


\section{Hybrid Methods}
Each recommendation method has its own virtues and drawbacks. 
This fact has led scholars to combine them in order to have a recommender system that benefits from those virtues and be able to overcome most of the drawbacks. Finally, researchers came up with the idea of using a hybrid method for recommender systems. A hybrid filtering method may use a combination of collaborative filtering with demographic filtering or collaborative filtering with content-based filtering to have boosted results. For instance, Balabanovic et al.\cite{balabanovic1997fab} created a recommender system named {\it Fab} which extracts user profile of interest on web pages by content filtering techniques and uses that information for collaborative filtering. Moreover, the hybrid method can involve different recommender systems based on the confidence that they have on predicted ratings or recommendations\cite{burke2002hybrid}. Additionally, different techniques from one method can be combined together to create a new hybrid RS. 

\begin{figure}[H]
\centering
\includegraphics[trim={0cm 0cm 0cm 0},clip, width=0.82\textwidth]{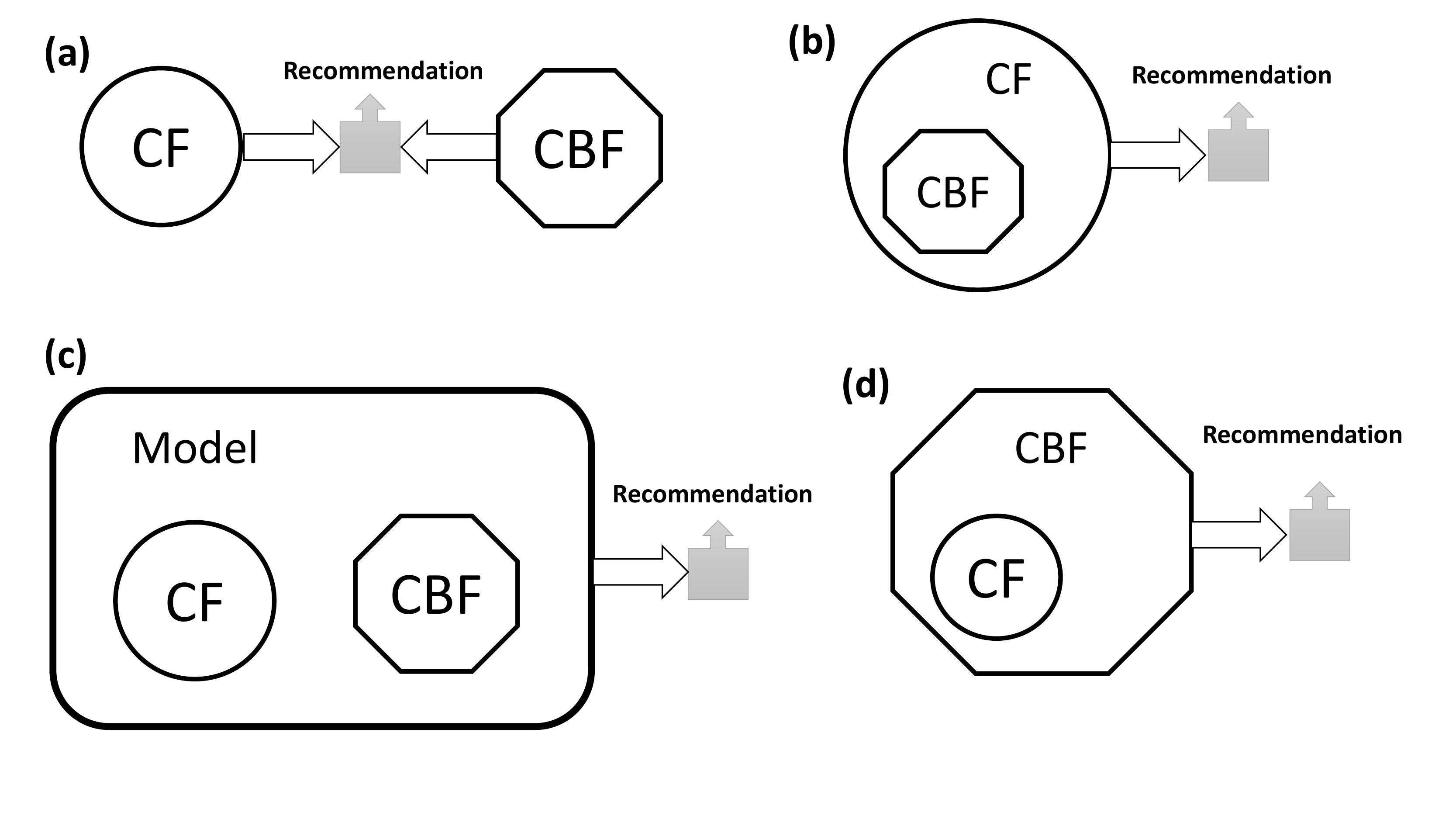}
\caption{Different methods of combining CF and CBF}
\label{fig:cfCBF}
\end{figure}

Predominantly, collaborative filtering has been combined with content-based filtering to make a hybrid method. Babodilla et al.\cite{bobadillaSurvey2013} categorized them in four different groups that is shown in figure~\ref{fig:cfCBF}. Figure~\ref{fig:cfCBF}a indicates the methods that combines CF and CBF with a weighting method. It may rank the items from both and recommend the top best items from them. Figure~\ref{fig:cfCBF}b shows the methods that use CBF methods to extract features and send it to CF to make the final recommendation. The example we mentioned from Balabanovic~\cite{balabanovic1997fab} used this technique. Furthermore, the prediction from CBF can be an input of CF as well. In figure~\ref{fig:cfCBF}c, a unified model is depicted  that utilizes CF and CBF to have their output for another classifier, such as rule based classifier or a probability model. Figure~\ref{fig:cfCBF}d depicts a model that uses output from CF for CBF. For example user ratings can help CBF characterize users better. 

\section{Evaluation Criteria}
Recommender systems should be evaluated for many reasons, such as comparison of quality of techniques. Evaluation metrics play a significant role in comparing several solutions for the same problem. Evaluation metrics are categorized in four main groups\cite{bobadillaSurvey2013}: (a) prediction metrics, such as accuracy (b) set recommendation metrics, such as Precision and Recall (c) rank recommendation metrics like half-life and (d) diversity and novelty. 

\subsection{Quality of the prediction}
Quality of prediction is the first criterion that has been used to compare recommender systems. One of the widely used prediction metrics is Mean Absolute Error (MAE) and other metrics derived from it such as mean square error, root mean square error (RMSE), and normalized mean absolute error. 

We define $U$ as the set of users of RS, and $I$ as the set of RS items. Then, $p_{u,i}$ is the prediction of item $i$ on user $u$, $r_{u,i}$ is the rating of user $u$ on item $i$ and ``$\bullet$" means user $u$ has not rated item $i$. Let $O_{u}=\left \{ i\in I \mid p_{u,i} \neq \bullet \wedge r_{u,i} \neq \bullet \right \}$ be the set of items rated by user $u$ with prediction values. The error is defined as the difference  between prediction and real value: $\left |  p_{u,i} -  r_{u,i} \right |$. Equations~\ref{eq:mea} and~\ref{eq:rmse} show how we should calculate the mean absolute error (MAE) and the root mean square error (RMSE) respectively.

\begin{equation}
MAE = \frac{1}{\# U }\sum_{u \in U}\left ( \frac{1}{\# O } \sum_{i \in O_{u}} {\left | p_{u,i} - r_{u,i} \right |} \right )
\label{eq:mea}
\end{equation}
where ``$\# \{*\}$" means the number of elements of the set $\{*\}$ or cardinality of it.
\begin{equation}
RMSE = \frac{1}{\# U }\sum_{u \in U}
\sqrt{ \frac{1}{\# O } \sum_{i \in O_{u}} {\left ( p_{u,i} - r_{u,i} \right )}^{2} }
\label{eq:rmse}
\end{equation}

Another metric is coverage which can be interpreted as the capacity of predicting from a particular metric\cite{ge2010beyond}. It calculates the percentage of situations in which at least one out of the $k$ neighbors of each active user rates an item that has not been rated yet by that active user\cite{bobadillaSurvey2013}. The total coverage of a recommender system equals to the average of all users' coverage. We define $K_{u,i}$ as the set of user $u \in U$ which have rated the item $i$, $C_{u}=\left \{ i\in I \mid r_{u,i} = \bullet \wedge K_{u,i} \neq \varnothing \right \}$ as the set of items that have not been rated by user $u$ and at least one of the neighbors rated it, and $D_{u}=\left \{ i\in I \mid r_{u,i} = \bullet \right \}$ as the set of items that have not been rated by user $u$. We have equation~\ref{eq:cu} to calculate coverage:

\begin{equation}
coverage = \frac{1}{\# U }\sum_{u \in U}
 \frac{1}{\# O } \sum_{i \in O_{u}} {\left ( 100 \times \frac{\#C_{u}}{\#D_{u} } \right )}
\label{eq:cu}
\end{equation}

\subsection{Quality of the set of recommendations}
For some users, having a reduced set of items is more important that having one item recommended. Precision, recall and F1 are the most important metrics to evaluate the quality of the set of recommendations. Precision indicates the rate of relevant recommended items to all of the recommended items. Recall is about the rate of relevant recommended items to all of the relevant items and F1 is a combination of precision and recall. We consider $X_{u}$ as the set of recommendations to user $u$, and $Z_{n}$ as the set of $n$ recommendations to user $u$, we calculate the aforementioned metrics by making $n$ test recommendation to user $u$. By considering a $\theta$ as threshold, we have equation~\ref{eq:prec} for Precision.

\begin{equation}
precision = \frac{1}{\# U }\sum_{u \in U}
 \frac{\#{\left\{ i \in Z_{u} \mid r_{u,i} \geqslant \theta \right \} }}{n} 
\label{eq:prec}
\end{equation}
Equation~\ref{eq:prec} sums over the number of recommendations and normalizes them. Moreover, equation~\ref{eq:recall} calculates Recall.

\begin{equation}
recall = \frac{1}{\# U }\sum_{u \in U}
 \frac{\#{\left\{ i \in Z_{u} \mid r_{u,i} \geqslant \theta \right \} }}{\#{\left\{ i \in Z_{u} \mid r_{u,i} \geqslant \theta \right \} } + \#{\left\{ i \in Z_{u}^{c} \mid r_{u,i} \geqslant \theta \right \} } } 
\label{eq:recall}
\end{equation}

\noindent where $\#{\left\{ i \in Z_{u}^{c} \mid r_{u,i} \geqslant \theta \right \} }$ denotes the number of relevant items that have not been recommended. 


\noindent The $F1$ measure is calculated in equation~\ref{eq:f1}.

\begin{equation}
recall = \frac{2 \times precision \times recall}{precision + recall}
\label{eq:f1}
\end{equation}

In figure~\ref{fig:recal_pre} we depict the role of recall and precision for evaluating recommender systems.
\begin{figure}[ht]
\centering
\includegraphics[trim={2cm 6cm 2cm 0},clip, width=0.75\textwidth]{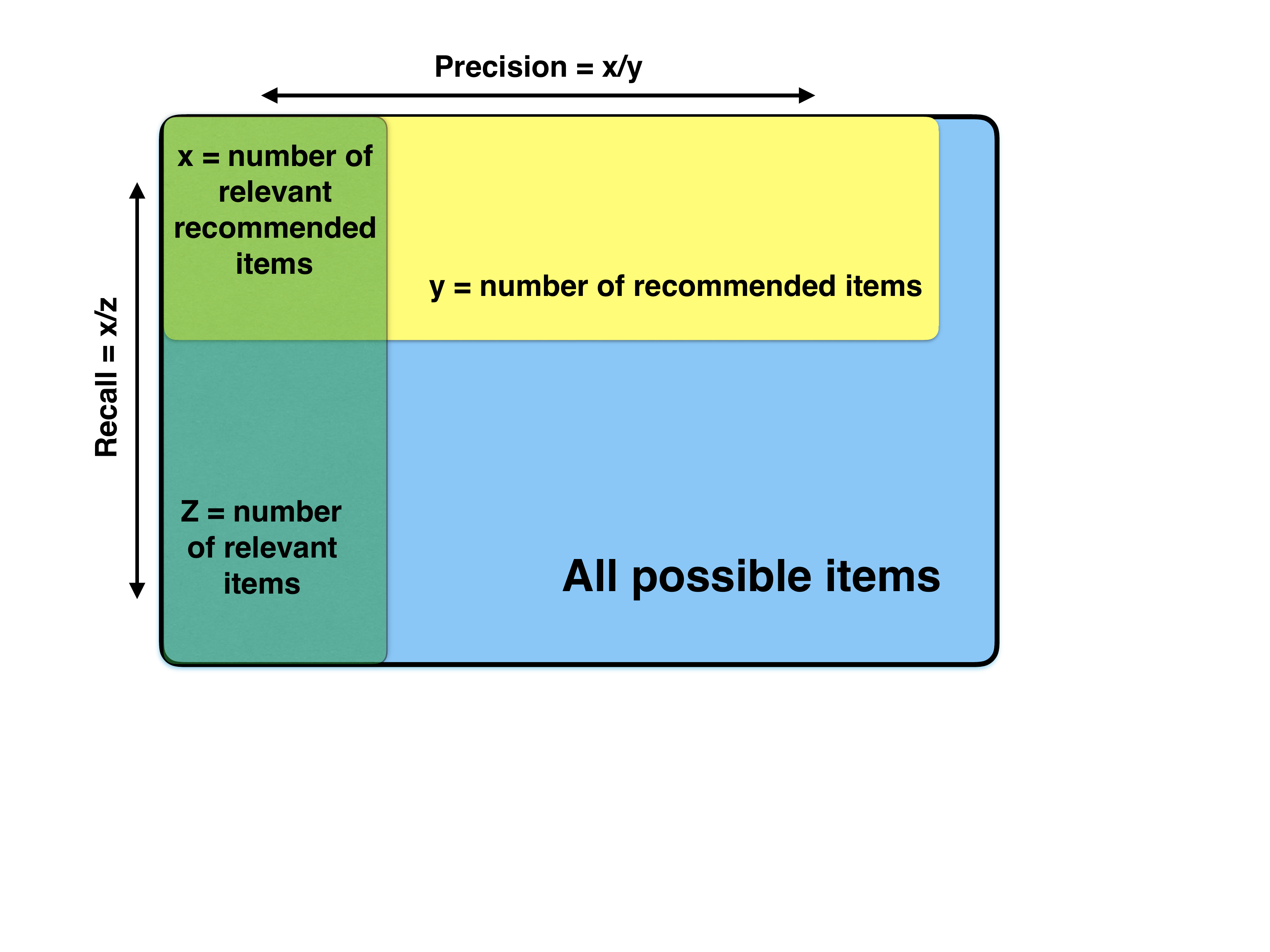}
\caption{Precision and recall with respect to all of the items.}
\label{fig:recal_pre}
\end{figure}

\subsection{Quality of the list of recommendations}
When we have a considerable number of recommendations, users give attention to the first items. Consequently, if the RS makes mistake in the first options, it is going to be a serious mistake. From information retrieval studies, we adopt the ranking measures that have been used in information retrieval can be applied here; {\it half-life} and {\it discounted cumulative gain} are the most popular measures for recommender systems. Equation~\ref{eq:HL} shows how we should calculate half-life. It assumes that users loses their interest of the following items in the list exponentially.

\begin{equation}
Hl = \frac{1}{\# U }\sum_{u \in U}\sum_{i=1}^{N}
 \frac{max(r_{u,p_{i}}-d,0)}
 {2^{(i-1)/(\alpha-1)}} 
\label{eq:HL}
\end{equation}
where $p_{1},...,p_{n}$ represents the recommendation list, $r_{u,p_{i}}$ is the true rating of user $u$ for item $p_{i}$, $d$ is the default rating and $\alpha$ is the number of items that have 50\% chance to be reviewed by user. 

Similar to half-time, discounted cumulative gain (DCG) considers a logarithmic decay in users' interest.

\begin{equation}
DCG = \frac{1}{\# U }\sum_{u \in U} {\left ( r_{u,p_{1}}+
\sum_{i=2}^{k}{\frac {r_{u,p_{1}} }{\log_{2}{(i)}}} \right )}
\label{eq:dcg}
\end{equation}

\noindent where $k$ is the rank of recommended items in the list of recommendations. 

\subsection{Novelty and diversity}
In some applications the RS needs to recommend novel items, because companies want to sell their new items as well. Further, some users may want to explore a new type of items. Therefore, there should be a metric to compare recommender systems based on this criterion. In this case, we want to know up to what extent a RS can recommend diverse items. Novelty and diversity are two main metrics that are useful here. There is no standard way to define these metrics and scholars tend to use different ways to calculate them. However, numerous authors used equation~\ref{eq:diversity} and~\ref{eq:novelty} to calculate diversity and novelty respectively\cite{huang2007comparison}:

\begin{equation}
diversity_{Z_{u}} = \frac{1}{\# Z (\# Z -1)}\sum_{i \in Z_{u}} \sum_{j \in Z_{u} , j \neq i} 
{\left [ 1-sim(i,j) \right ]}
\label{eq:diversity}
\end{equation}

\noindent where $sim(i,j)$ denotes item to item collaborative filtering similarity measures, and $Z_{u}$ is the set of $n$ recommendations to user $u$. In equation~\ref{eq:diversity}, diversity is calculated by summing over the similarity between pairs of recommended items and normalizing it.

\begin{equation}
novelty_{i} = \frac{1}{\# Z -1} \sum_{j \in Z_{u}} 
{\left [ 1-sim(i,j) \right ]}, \quad i \in Z_{u}
\label{eq:novelty}
\end{equation}

Equation~\ref{eq:novelty} shows how to obtain novelty for each recommended item. It returns the normalized similarity between item $i$ and all other recommended items in $Z_{u}$. Note that, sometimes, novelty is vital, because there are some items which most of the users do not buy frequently (like refrigerator). Thus, if a user buy one of them, most likely he or she will not buy it again in the near future. Then, the RS should not continue to recommend it to the user. However, if the user tries to buy them again, the RS should learn that and include them in the set of recommended items. There are some other metrics that might or might not be important for an RS designer. For example, stability in the RS prediction. It implies that the set of recommendations should not be changed drastically through the time\cite{adomavicius2012stability}.

\chapter{Context-aware Recommender Systems}
\label{ch:context}
A wide range of recommender system techniques concentrate on the most relevant item based on user ratings. However, there is other useful information that can be collected in order to help the recommender system. This information may consist of time, place, job or any other beneficial information about the user or a group of users. As a result, in addition to the two traditional components of a recommender system, i.e. item and user, we have other information as well. This information is referred to “ contextual information” and can be  applied in special circumstances. For example, information about time can help us in recommending a travel package, or a web page. 
Additionally, mobile recommender systems attract attention, because a significant number of users have mobile devices and information such as location and time can be extracted from those devices in order to help the recommender system to understand the context better.

This topic leads us to a wider area of information that should be taken into account concerns user behavior in different circumstances, because a recommender system with more contextual information can be more accurate. For instance, a music recommender can be more accurate if it considers places of interest, in-car music, music while reading, and even the mood of the listener\cite{Adomavicius2015contextCh}. Another example is the Netflix recommender system that uses locational contextual variables such as city or zip code and time to provide context specific recommendations. Reed Hasting, the CEO of Netflix, claimed that they can improve the performance of their recommender system up to 3\% when considering such contextual information\footnote{Watch the video at 44:40 minute https://www.youtube.com/watch?v=8FJ5DBLSFe4}. In general, context-aware recommender systems consist of three main parts, pre-filtering, post filtering, and modeling, which we explain in more detail in the following sections. 

\section{Context in Recommender Systems}
Before we talk about context in recommender systems, we should know what the context is in general. The definition of context in Webster's dictionary is: ``the parts of a discourse that surround a word or passage and can throw light on its meaning; condition or circumstances which effect something; the interrelated conditions in which something exists or occurs :  environment, setting the historical context of the war"\footnote{https://www.merriam-webster.com/dictionary/context?}. As can be seen, the definition is not precise, and that suggests that the concept of context is a multidisciplinary concept that can have a different definition in each field of study. In computer sciences, and specifically in recommender system studies, a context is the information that can improve the performance of the system and cannot be measured just by tracking user rating or item rates. 

The traditional methods, particularly collaborative and content-based filtering, use two important fundamental elements of a recommender system, i.e. item and user, to predict the ratings. Therefore, we can assume that a recommender system is a function that takes users and items and returns ratings:

\begin{center}
$R: User \times Item \rightarrow Rating$
\end{center}
 \noindent In this function, the input is actually in 2 dimensions since it only considers users and items. However, when we add the concept of context in our recommender system, it becomes another input parameter to the rating function. Then we have:
 \begin{center}
$R: User \times Item \times Context \rightarrow Rating$
\end{center}

Context can be considered as a vector that contains different contextual information.
There have been two main representational approaches for context: hierarchical and tensor representation. Hierarchical representation is introduced by Palmisano et al.\cite{palmisano2008usingContext} suggesting granular information as contextual dimensions. In their model, contextual information is defined as a set of contextual dimension {\bf K}, so that each contextual dimension $\it k$ is a set of $q$ attributes $k= \{k^{1}, ..., k^q\}$ and these attributes have a hierarchical structure to capture different types of context. $k^{q}$ is the finer or more granular level of information, while the $k^1$ defines the coarser or less granular level of contextual information. 

\begin{figure}[ht]
\includegraphics[width=.8\textwidth]{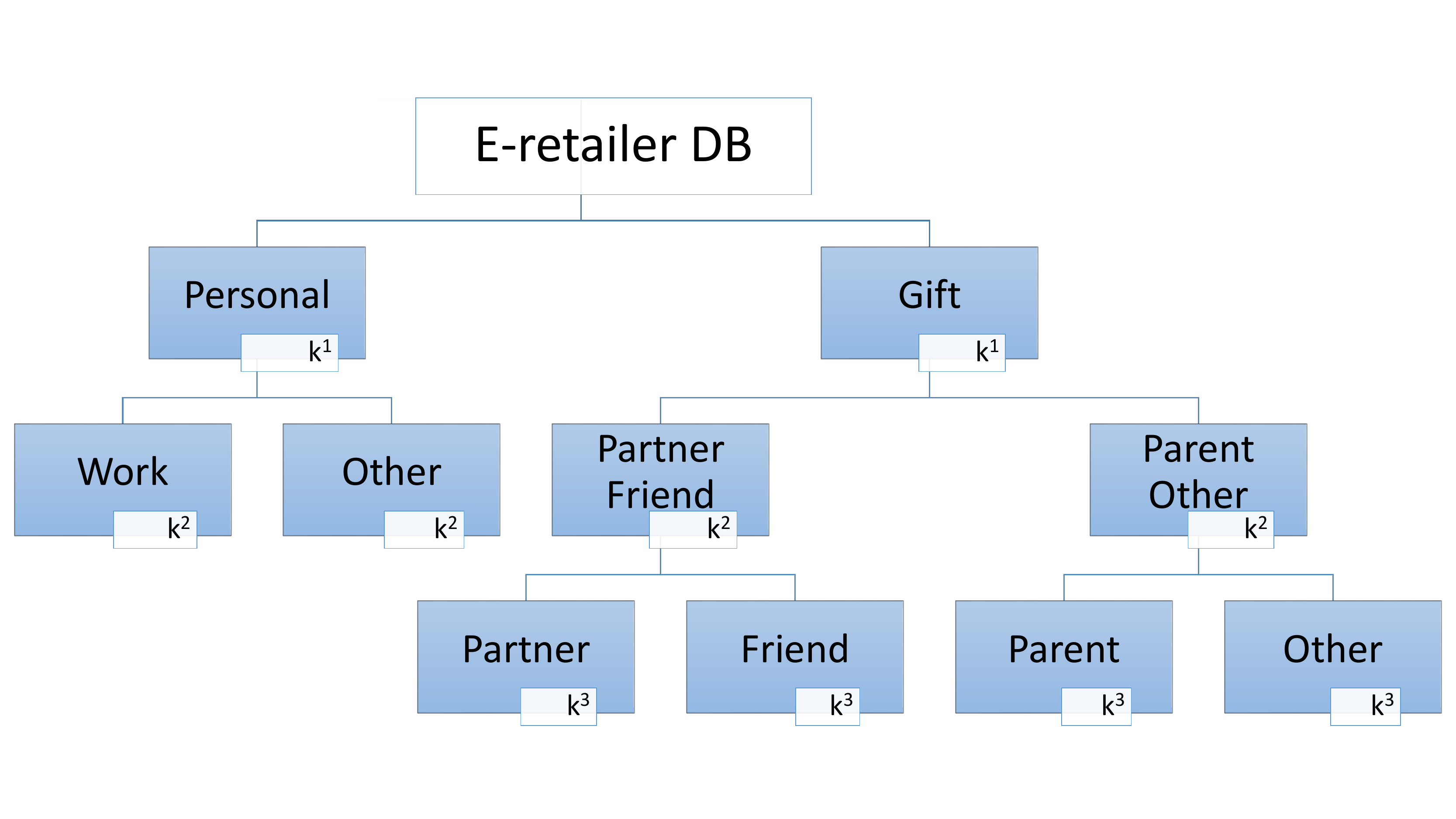}
\caption{contextual information represented by hierarchical structure.}
\label{fig:hierarch}
\end{figure}

As can be seen in figure~\ref{fig:hierarch}, which is an example from\cite{Adomavicius2015contextCh}, the root contains the coarsest level of information (all of the database). Then the next level is the information about whether the merchandise is for personal use or a gift; thus, we have $k^{1}=\{ Personal, Gift \}$. The next finer level of the hierarchy could be the values of either  ``Personal'' or ``Gift''. Subsequently, we have the next finer level $k^{2}$=\{PersonalWork, PersonalOther, GiftPartner/Friend, GiftParent/Other\}. 

A different way to represent context is the way that mathematicians work with tensors. If we let $D_1, D_2, \dots, D_n$ be dimensions the input vector for the $R$ function:
\begin{center}
$R: D_1 \times D_2 \times \dots \times D_n \rightarrow Rating$
\end{center}
Two of these dimensions are item and user, and the rest are contexts. In the tensor representation method, each dimension $D_i$ is a Cartesian product of some attributes $A_{ij}, (j=1,2,\dots,k_i)$; that is, $D_i  \subseteq A_i1 \times A_i2 \times \dots A_{ik_i}$. For more illustrations, we can consider a recommender system in which a user has the information such as $UName$ and $Address$ and $Age$. This can then be shown as $User \subseteq UName \times Address \times Age$; likewise, the item dimension could be  $Item \subseteq IName \times Type \times Price$, and if we consider the $Time$ as our context, it could be $Time \subseteq Year \times Month \times Day$. Figure~\ref{fig:recal_pre} shows this tensor model for the stated example. 

\begin{figure}[ht]
\centering
\includegraphics[trim={1cm 6cm 1cm 0},clip, width=0.75\textwidth]{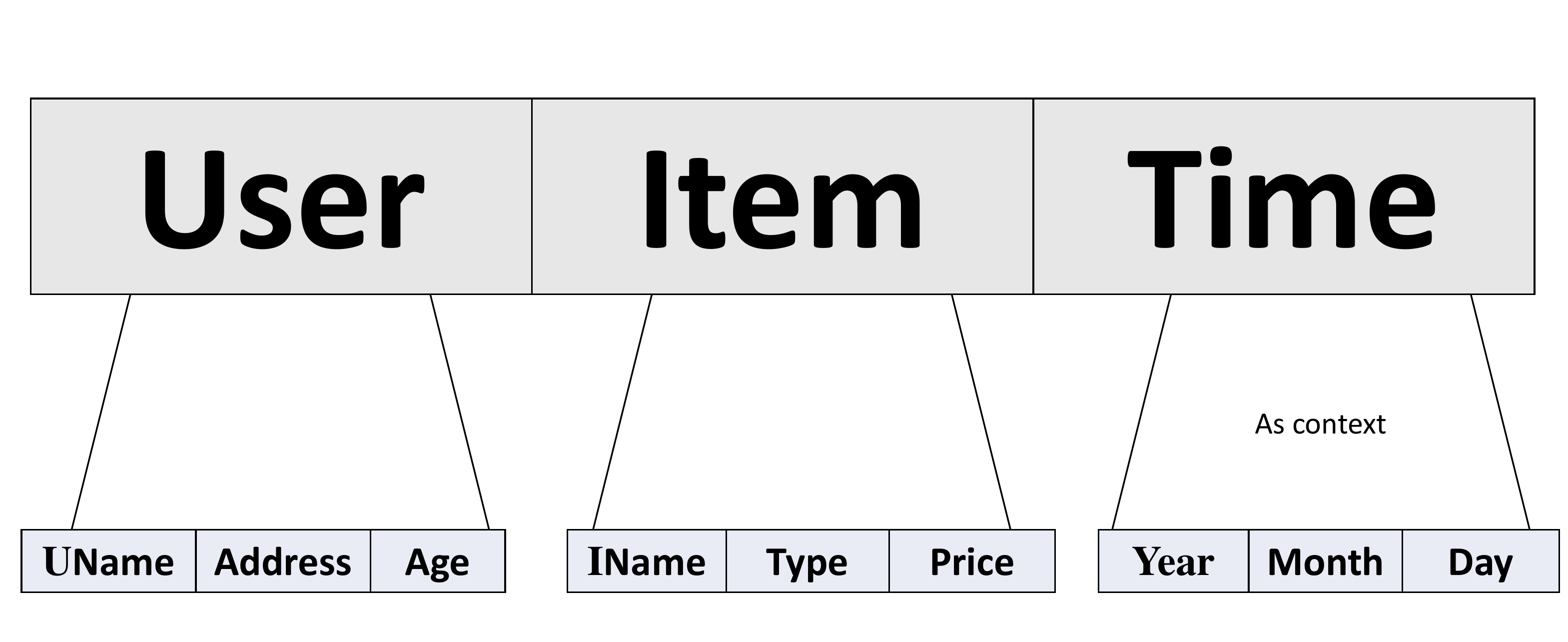}
\caption{Precision and recall with respect to all of the items.}
\label{fig:recal_pre}
\end{figure}

\section{Obtaining contextual information}
Before working on a recommender system that uses context, we need to know how the context is going to be obtained, albeit there are some context-aware recommender systems (CARS) that assume the contextual dimensions have already been provided. The availability of information about the items, users and some other circumstances related to them or the interaction between them plays a significant role in content acquisition. In general, there are three ways to obtain context\cite{Adomavicius2015contextCh}:
\begin{itemize}
\item {\bf Explicitly} --- As we mentioned in Chapter~\ref{ch:methods}, in the explicit approach, the information is gained directly from entities. For example, a website or a company may provide a survey for users and ask them to fill it in. Likewise, the information of location or time can be extracted from the users' device.

\item {\bf Implicitly} --- This type of information needs a monitoring system to observe the users and interactions. It should be noted that the source of information is accessed directly. For example, frequent changes in the GPS of a user extracted from her device is implicit information about the user that suggests the user may not stay for a long time in a specific location.

\item {\bf Inferring} --- In this approach, the recommender system should infer information from other data that has already been extracted. The information here is hidden and requires special algorithms to be revealed. For example, if a machine learning method recognizes the type of person who is watching TV at home, it can help the recommender system to recommend better TV shows which are more desirable for that specific person. 
\end{itemize}

All of the aforementioned methods should be performed as part of the data collection process, because the recommender system relies on the data and predicts rates based on them. Another important issue here is the relevance of the extracted context. For example, a book store (either online or traditional) can capture information from a buyer regarding their purpose in buying the book, the planned reading time, and general information about the stock market at the time of buying. However, the information about stock market may not be applicable at all. Hence, the relevance of the information is important, and it becomes crucial in context-aware recommender systems because they work with larger databases than usual. Here it is necessary to have an expert in the domain of the application. Another example is a mobile recommender system that needs physical context such as time and position, and social context regarding whether the user is alone or not; interaction media context such as the type of device is also important context. In the case of technology enhanced learning (TEL), computing context, user context, and physical context are all important\cite{verbert2012context}. 

Besides using an expert or a manual approach to define the essential relevance context, there are some machine learning and data mining algorithms that help us to detect contexts automatically\cite{Odic2013}. Adomavicius et. al. in\cite{adomavicius2005toward} suggested that an expert should suggest some contextual features as candidate; then, by statistical methods, the most relevant one is extracted. For example, they did a pairwise t-test among candidate features. Another common way to assess the relevance of a context is stated by Baltrunas et al.\cite{baltrunas2012context} which it is suggested that some hypothetical contextual preferences should be offered to users as a survey. Then they ask users to respond to survey, and in this way they collect useful contextual information. They show that their system outperforms a recommender system that does not use context.

\section{Utilizing Context in Recommender Systems}

In order to utilize context in recommender systems, we should take into account two major approaches to using this information: {\it (i) context-driven querying} and {\it (ii) contextual preference elicitation and estimation}. The context-driven approach suggests that the recommender system should rely only on contextual information and try to relate the items and users based on the contextual information. Some scholars use it to create a mobile tourist recommender system\cite{cena2006integrating}.

\begin{figure}[ht]
\centering
\includegraphics[trim={0cm 0cm 0cm 0},clip, width=0.85\textwidth]{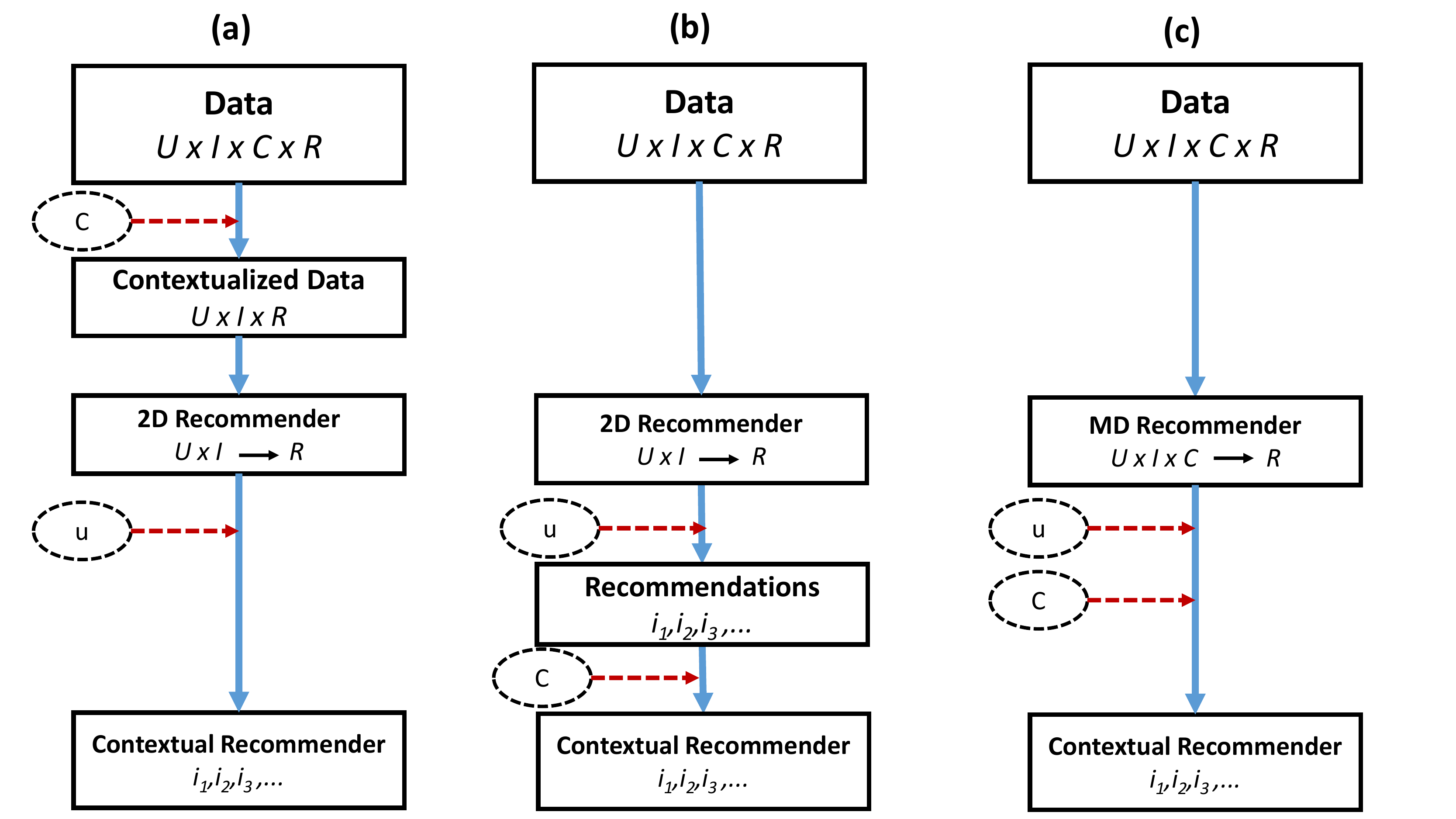}
\caption{Paradigms of using context in recommender systems. (a) Pre-filtering (b) post-filotering (c) contextual modeling.}
\label{fig:contextual}
\end{figure}

On the other hand, the contextual preference elicitation and estimation methods have engaged more context-aware recommender system researchers. Unlike the previous method, this one encourages us to learn the context and reinforce the collaborative or content-based filtering by using it. It should be noted that it is possible to design a recommender system that uses a combination of both general methods. We may recall that recommender systems are created based on partial user preferences (i.e. some ratings from some users), and the input record of recommended systems are a subset of $<user,item,rating>$. In context-aware recommender systems we have a new element known as ``context'' that changes the records to a new tuple which is $<user,item,context,rating>$. Now it is important to decide how and where in the recommender system we should use contextual features. Generally, we can use context either before selecting data records, after selecting them, or in the recommending process. Figure~\ref{fig:contextual} illustrates these methods, and we will explain them in more detail in the following sections. 

\subsection{Contextual Pre-filtering}

In this recommendation paradigm (figure~\ref{fig:contextual}a), the information about a certain context $c$ is used to select or filter relevant data; then it is fed to the conventional $2D$ (i.e. $User \times Item$) methods such as collaborative or content-based filtering. For instance, context $c$ is considered as a query to find relevant ratings data\cite{Adomavicius2015contextCh}. For a more detailed illustration, if we assume that a viewer wants to watch a movie on Saturday, the recommender system first picks all the Saturday movies' ratings and feeds them to a collaborative filter to find the closest user, and then recommends the best items for the viewer. This method is called {\it exact pre-filtering}, because the data filtering is based on an exact value of a context. It can be seen that we turned a $3D$ input recommendation problem into a $2D$ one; after all, the collaborative filtering part of the recommender system does not deal with context anymore. It can be easily implemented by a selection and a projection over the database as following:
\begin{center}
{\small $\forall (u,i,c) \in U \times I \times T, R_{user \times item \times context}^{D} (u,i,t) = R_{user \times item}^{D[Context=c](User,Item,Rating)} (u,i)
$}
\end{center}

The downside of this method is the narrow context that it returns. For example, in a case where the context is $c=(Partner,Theater,Saturday)$, the recommender may not find the best movie that is playing in a good theater on Saturday that is good for the viewer to watch with a partner. In order to avoid this problem, Adomavicius et al.\cite{adomavicius2005toward} suggest using {\it generalized pre-filtering} which uses aggregated information and tries to generalize the contextual information. If we recall from our previous example, the recommender system could aggregate Saturday and Sunday together and show them by a new aggregated value named ``Weekend'', in which case it would find more options. We let $S_c$ be a segment of data that aggregates context, i.e. $c \in S_c$; then, in this method, the selection and projection steps would be:

\begin{center}
$R_{user \times item \times context}^{D} (u,i,t) = R_{user \times item}^{D[Context \in S_c](User,Item,Aggregate(Rating))} (u,i)
$
\end{center}

Moreover, it is even possible to use more than one aggregated context and to filter the data based on them. 

The aggregation reformation brings another problem, which is the need to find a ``right'' level of granularity. One may think about using an expert person for that, but to have an adaptive system for big data we need a computational automatic approach. In\cite{jiang2009improving}, the authors investigated different levels of generalization and compared the prediction accuracy of the recommender system in order to find the best level of generalization.

Another issue is the locality problem that happens because the recommendations come from the data that is pre-filtered by aggregated context from a specific segment of contextual information. For example, if you want to go out and have fun by watching a movie in a theater, then a pre-filtering context-aware recommender system may generate better recommendations for you. However, if you are at home and want to watch a movie on your TV, then it might be better to have a simple $2D$ recommender system. 

\subsection{Contextual Post-filtering}
In this method (figure~\ref{fig:contextual}b), contextual information is not considered until the last step of the recommendation. It means that the system takes the whole $3D$ database and makes decisions on this data; then, at the end, right before making the final list of recommended items, the contextual information is applied to adjust the final list. There are two main approaches to modify the final list based on the contextual information: {\it Filter out} the irrelevant items or {\it reorder} the items in the recommended list. Furthermore, the post-filtering technique is classified into heuristic (memory-based) and model-based ones.

In the heuristic method, the post-filter part of the recommender system searches for common item features for a given context and adjusts the list based on their quantity. For example, if you like some movies with specific actors, it will adjust the recommendation list to include more of those actors. This adjustment can be done by filtering (dumping) out the movies that do not have a specific number of those actors, or it can be accomplished via ranking the movies in the list based on the number of desired actors involved in them. 

In the model-based method, the post-filter can learn the probability of the popularity of a movie based on the its context. For instance, it may learn the likelihood of choosing a movie with a certain director. Then it uses that probability to adjust the recommendation list. This adjustment operation could filter out the items which have the relevance probability less than a certain threshold. Similar to the heuristic model, it can also rank the final list by weighting the items in it using the calculated probability. Panniello et al.\cite{panniello2009experimental} compare post-filtering and pre-filtering methods on two databases of an e-commerce and Amazon\footnote{Their dataset consists of some items purchased by students containing contextual information}. Their results suggest that weighted post-filtering performs better than the pre-filtering method, and the pre-filtering outperforms the filter post-filtering.

\subsection{Contextual Modeling}
In contextual modeling (figure~\ref{fig:contextual}c), the contextual information is used in the process of finding the unknown ratings. One common method is to deploy the context directly in the process of user rating prediction. In contradiction to the pre-filtering and post-filtering methods, this method uses the $3D$ recommendation function. That means that it operates like $Rating=R(User,Item,Context)$ where $R$ is a prediction function that predicts each user's rating on a target item. A similarity function can be used to find the similarity between the $<user,item,context>$ tuples. The unknown ratings are predicted with respect to those tuples that have rates on items. Moreover, the ratings involved in this calculation are inversely related to the similarity metrics. Equation~\ref{eq:Ruic} shows the a prediction method for an unknown rate $r_{u,i,c}$ for $<u,i,c>$ which is a tuple in the database:

\begin{equation}
r_{u,i,c}=k\sum_{({u}',{i}',{c}') \neq (u,i,c)}
{W\left (({u}',{i}',{c}') , (u,i,c) \right )} \times r_{{u}',{i}',{c}'}
\label{eq:Ruic}
\end{equation}

\noindent where $k$ in a normalization factor, and $W\left (({u}',{i}',{c}') , (u,i,c) \right )$ is the ``weight'' of the rating $ r_{{u}',{i}',{c}'}$ participating in calculating the prediction rate which can be the inverse of the Euclidean distance between  $({u}',{i}',{c}')$ and $(u,i,c)$. In other research\cite{Adomavicius2005} the aggregated information of the context shows better performance; furthermore, the authors consider the distance equal to zero wherever the the context in two tuples is not the same (i.e. if $c \neq {c}'$ then $dist\left (({u}',{i}',{c}') , (u,i,c) \right ) = 0$). 

Additionally, Oku et al.\cite{oku2006context} use the additional context in the $3D$ database and use the support vector machine (SVM) classification, which looks into the items and corresponding ratings as two sets of ``like'' and ``dislike'' and creates the hyperplane based on the support vectors, then recommending the items that fall on the like side of the hyperplane.

\afterpage{\null\newpage}
\chapter{Social-based Recommender Systems}
\label{ch:socialInfo}
The emergence of social networks and their drastic growth suggests that the tremendous information within them could be helpful in many applications including recommender systems. Moreover, the overload of resources (i.e. items and data in general) makes the process of making decisions even harder for social media users. Therefore, we need a social media-based system that channels the resources in social media. For example, by learning from the new types of data extracted from online social networks such as tags and relationships we can help a recommender system to find similar users in a better way.

In general, social information is useful for three main reasons\cite{bobadillaSurvey2013}. First it can be deployed to improve the quality of prediction. For example, the RS may infer that since two users are friends in a social network, it is possible for them to have the same taste for items. This can help collaborative filtering methods. In\cite{Woerndl2007PhysicalSocial} the authors show that social information enhances the result of collaborative filtering. Second, it can even be used to create a new recommender system. Here, the goal is not to improve a pre-designed recommender system but to propose a new way to generate an RS based on social information. Siersdorfer and Sergei in\cite{Siersdorfer2009SocialWeb} used the multi-dimensional social environment of a specific user to create a social recommender that suggests users, items or groups to that specific user. The third purpose of social filtering is just to analyze the relationships between social information and collaborative entities. For example, correlation between recommender and recommendee may be important for decision-making problems.

Initially it was thought that social information could be used to create a trust network for recommender systems, but weak generalization led the scholars to have a wider overview on information from social networks\cite{sun2015recommender}. Moreover, some researchers believe that content recommendation is an important subject that should be considered in social-based recommender systems\cite{guy2015social_ch}. 

\section{Recommendation related to contents}
In social media, content plays a vital role, whether it is going to be recommended to users to use, or to generate new content. For example, a social recommender for Facebook users may recommend news or video to the users to read or watch. On the other hand, it may recommend topics based on the trends or tags to users to post a text about it. The content could be the comments of a user, the tags used, or the votes or ratings (e.g. like and dislike). These contents in addition to the relations among users, can give us an invaluable opportunity to have a more effective recommender system. Golbeck\cite{Golbeck2006} uses membership forms from the ``FilmTrust" system, which is a web-based social network and has a movie rating and review system. The author uses trust between individuals as the weight of their mutual rating on an item, then estimates the unknown rating based on the weighted known ratings. Her results show that this information can improve movie recommendations. 

Guy\cite{guy2015social_ch} mentions the important ``key domains'' in social recommender systems. The first important one is the {\it blog} which is one of the classic social media. A blog owner (a person or company) writes about a topic on the blog and it creates a blog post. The owner or users of the blog can add posts about the topic or interpret it in the comments. The blog itself can be an item to recommend. Moreover, the content of the posts and the reactions to them can be considered as a context or extra information to help the recommender system.

The next key content is {\it multimedia}, which is quite challenging since extracting the actual content in an audio file or a video is computationally expensive, and it returns an enormous amount of content. The most famous social media for multimedia purposes is YouTube.  Davidson et al.\cite{davidson2010youtube} use co-visited video counting and associated rule planning\cite{zhang2002association} to predict the score of a video. They suggest that the YouTube recommender system should recommend fresh and diverse videos with respect to the video that the user has recently watched or reacted to. Beyond that, they state that the user should understand why a video was recommended to them.

{\it Question and answer} is the next important content in special Q\&A websites like StackOverFlow and Yahoo Answers. The main issue here is to recommend other relevant questions and/or appropriate answers. Another content related to online social network is {\it news}. Social news broadcasters such as Digg, Reddit, or Google Readerlet try their best to recommend the most relevant and popular news to the readers.
Research from Google\cite{liu2010personalized} creates a distribution of user clicks over a year tallied for each month. Then they use this information for computing the distance and then similarity to feed to the collaborative filtering part of their recommender system. They improved the pure collaborative filtering method by 7\% via this technique. It should be noted that recommending the freshest and most recent post is extremely crucial in both {\it question and answer} and {\it news} social networks.

\begin{figure}[ht]
\centering
\includegraphics[trim={1.5cm 1cm 1.5cm 0cm},clip, width=0.95\textwidth]{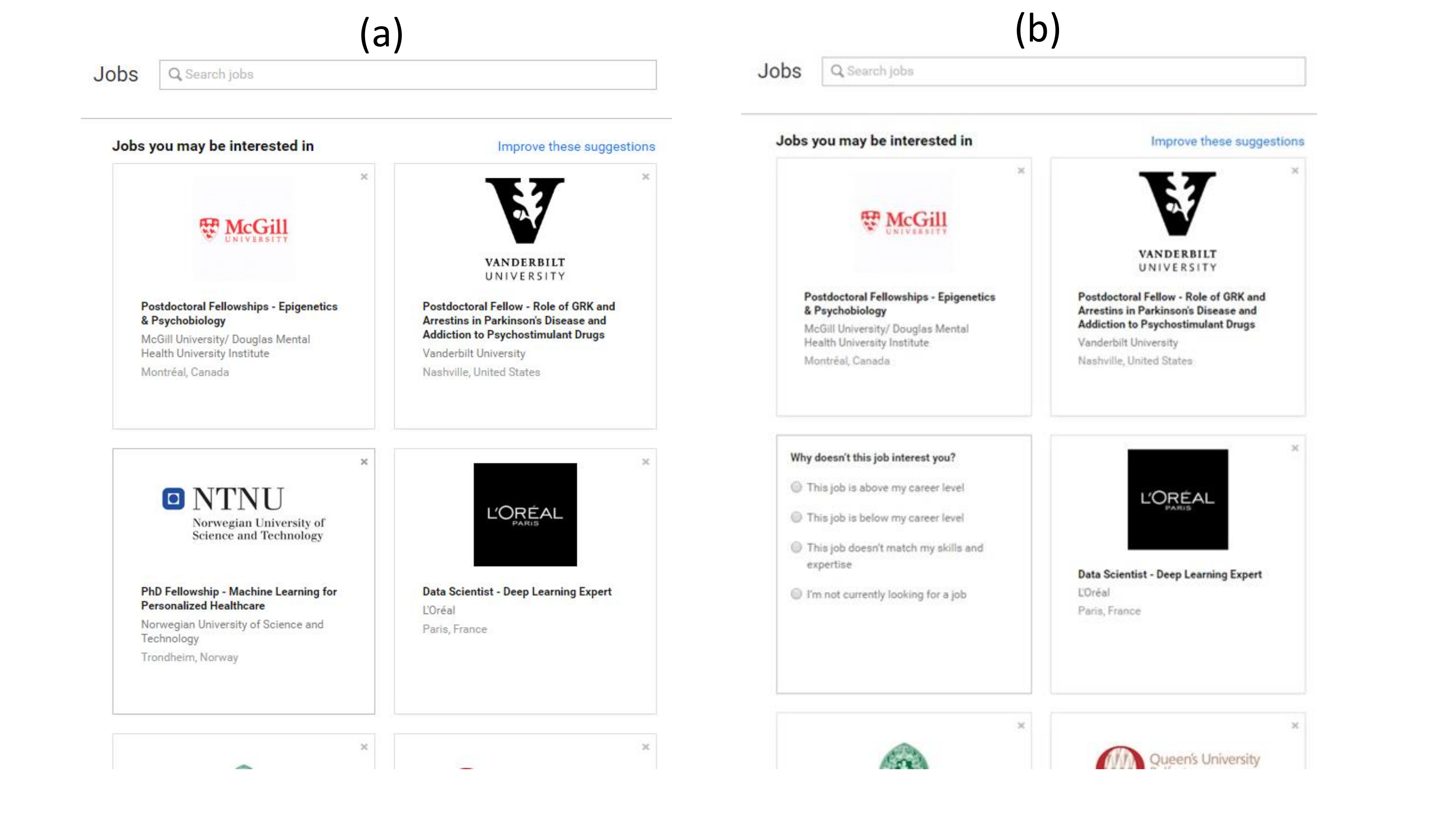}
\caption{Example of job recommendation by ResearchGate (a) Recommend opportunities (b) User can delete an opportunity from the recommended list and leave feedback.}
\label{fig:rgate}
\end{figure}

The Other content is about `{\it jobs}. LinkedIn is the best known website one in this area. Additionally, ResearchGate is recommending jobs and opportunity. The significant impact of this subject on people's lives make it an attractive one for recommender system scholars. Figure~\ref{fig:rgate} shows a profile on ResearchGate that recommends some job opportunities based on the user profile; the user can interactively purify the recommendations by leaving feedback on the recommended option. 

{\it Microblogs} are among the other contents that have become popular with Twitter. Here, the concept of ``follower'' and ``followee'' can help us to recommend tweets or people to users to follow. Most of algorithms use one of the following aspects on Twitter to recommend content: candidate selection, topic relevance or social voting. Social voting is about the number of user's followees that follow the user and also follows the posts that the users tweeted or followed. Comprehensive research\cite{Chen:2010:STE:1753326.1753503} on different algorithms of recommending URL on Twitter shows that social voting works better than topic relevance.

\subsection*{Social information in movie domain recommendation}
Carrer-Neto et al.\cite{carrer2012social} use semantic knowledge extracted from movie meta data along with the data extracted from the profile of the user. They assume that in the database they have, the user defined her ``social aperture'' by choosing one of these options: ``Moderate'', ``Liberal'' or ``Conservative''. 

If the user is moderate, they use 25\% of her friends' ratings to calculate her rating. If the user is liberal, both her own ratings and her friends' ratings will be considered equally and if she is conservative, then they only use the ratings calculated for the user. Their results show that using this social information outperforms the case where they did not use it.

\section{People Recommendation}
One of the main duty of a social recommender system is recommending people to each other. Social media websites must use an algorithm to suggest relevant or similar people to each other. Guy\cite{guy2015social_ch} discusses that relation between people on social networks has different dimensions. For instance, in Facebook, you and another user may become friends to each other, but in Twitter you may follow a user without having followed back by that user. Here we have the issue of ``symmetric'' versus ``asymmetric" relations. Moreover, in this example, you need to send an invitation to become somebody's friend, but on Twitter you can follow a user without her ``confirmation". Thus, you may face different social network either with or without confirmation. Sometimes a relation is a temporary one just to organize a meeting or an event. But in other cases, your relation is permanent. For example you may be in an online group of researchers from you laboratory. This indirect relation is considered permanent. 

An example of recommending people has been studied by Geil et al.\cite{geil2014wtf}. They use the ``Who To Follow" (WTF) algorithm on GPU. The algorithm first finds the circle of trust (CoT) of the user, which is the 1000 nodes closest to the user, and then creates a bipartite graph of individuals from the CoT on one side and the ones they follow on the other side. Then it uses Twitter's Money algorithm and assigns a similarity and relevance number to all nodes, after which it propagates the similarity value to followees and the relevance to followers. Finally, it recommends people with the highest relevance scores. 

\section{Group Recommendation}
Another issue in a social recommender system is ``group recommendation". It is important to determine that whether the recommender system is going to recommend items to a group or only to an individual. For example, in a case where the recommender system is going to recommend a TV show, if a group like a family wants to watch it, the system may recommend different items in comparison to the situation in which it deals with just one person. Consequently, some other questions that matter are what kinds of groups we are going to make our recommendation to, and how similar the members of the groups are to each other. One application here could be recommending some music to a group of people who are working out in a gym. Profile aggregation and recommendation aggregation are the most common approaches in this field.

In order to aggregate the rating of a group, we need to consider the type of strategies used to obtain the group rating. For this purpose, group recommending take into account three main strategies\cite{gartrell2010enhancing}:
\begin{itemize}
\item {\it Average satisfaction} which assumes equal importance for all the members of the group. Let $GR_i$ be the group rating on the item $i$, then in order to calculate it, we use equation~\ref{eq:ave} that simply calculates the average.
\begin{equation}
GR_i=average(r_{u,i})=\frac{\sum_{u=1}^{n} {r_{u,i}}}
{n}
\label{eq:ave}
\end{equation}
where $r_{u,i}$ is the rating of user $u$ on item $i$ and $n$ is the number of members in the group.
\item {\it Minimum misery} is used when we want to give special attention to the members of the group that rate an item very low. In this case, the group average is the minimum rating of all members, i.e. $GR_i=min\{r_{u,i}\}$.

\item {\it Maximum satisfaction} which is concerned with the members that rate an item higher than other members of the group. Then the group rating is the maximum of the rating of the members, i.e. $GR_i=max\{r_{u,i}\}$.
\end{itemize}

Nevertheless, the aforementioned strategies are not accurate enough to describe the aggregated group rating on an item. Hence, Gartell et al.\cite{gartrell2010enhancing} try to use social information in order to have a better group descriptor. They define a social weight $w_{u,v}$ as the contact frequency over a specific time. We can generalize it to a proportion of the number of tags two connected users have in common. Equation~\ref{eq:sg} obtains this descriptor:
\begin{equation}
S(G)=\frac{2 \cdot \sum_{u,v \in G}  {w_{u,v}} }
{|G| \cdot \left (|G|-1 \right )}
\label{eq:sg}
\end{equation}
Then they use it to define how much they should rely on min, max or average rating of the group. Basically, they say that if the social descriptor is not high or low, the average is desired, but if it is high or low, the maximum or minimum should be considered accordingly.

\section{Immediate Friend Inference}
If the access to social data about users is provided, we can involve the friends of a user to recommend the best suitable item to her. The impact of immediate friends (i.e. the friends with one hop distance) and a probability-based inference is discussed in\cite{he2010social} by He and Chu. They assume that the ratings are integers; then, they try to find out what is the probability of rating of user $u$ on the item $i$, i.e. $R_{u,i}$, given the set of attributes $a_{u}$ of user, set of attributes $b_i$ of item and the rating of the neighbors, i.e. $R_{v,i}$, for that item. They use naive Bayesian assumption and reach the equation~\ref{eq:condition}.

\begin{dmath}
P\left (   R_{u,i}=k | B=b_i , A=a_u, \{  R_{v,i}=r_{v,i}: \forall v \in U_i \cap N_u \}
\right) \\
= \frac{1}{Z} P\left (   R_{u,i}=k | B=b_i\right) \times P\left ( R_{u,i}=k | A=a_u\right) \\
\times P\left (   R_{u,i}=k | \{  R_{v,i}=r_{v,i}: \forall v \in U_i \cap N_u \}
\right)
\label{eq:condition}
\end{dmath}

\noindent where $B$ is the random variable standing for the set attribute of item $i$, $b_i$ is the set of values of attribute of item $i$, $A_u$ is the random variable standing for the set attribute of user $u$, $a_u$ is the set of values of attributes of user $u$, $v$ is a neighbor of user $u$, $U_i$ is the set of users that rated item $i$, $N_u$ is the set of the neighbors (friends) of user $u$, and $R_{v,i}$ is the ratings of neighbors of user $u$ on item $i$. 

Now we need to calculate each probability independently. The probability of the rating of user $u$ given set of attributes for item $i$ is the user preference. This means that in order to calculate the user preference we should find the probability $ P\left (   R_{u,i}=k | B=b_i\right)$; with the naive Bayesian assumption we have equation~\ref{eq:user_pref}.

\begin{dmath}
P\left (R_u=k | B=b_i \right)
= \frac{ P(R_u=k) \times P(B_1,B_2, \dots ,B_n | R_u=k)}
{P(B_1,B_2, \dots ,B_n)} \\
= \frac{ P(R_u=k) \times \prod_{j=1}^{j=n}  P(B_j | R_u=k)}
{P(B_1,B_2, \dots ,B_n)} , B_j \in \{B_1,B_2, \dots ,B_n \}
\label{eq:user_pref}
\end{dmath}

\noindent where $P(R_u=k)$ is the prior probability that the user $u$ gives a rating $k$, and $P(B_j | R_u=k)$ is the conditional probability that each item with attribute $B_j$ in $B$ gets the value $b_j$ given $u$ rated it with $k$. For example, $P(actor= Al Pacino | R_u=5) = 0.9$ means that the probability that Al Pacino plays in a movie given the movie received the rate 5 equals to 0.9. Equation~\ref{eq:R_K} and equation~\ref{eq:R_bj} calculate the two nominator probabilities in the previous equation by a simple counting over the database.

\begin{dmath}
P(R_u=k)=\frac{|I(R_u=k)|+1}
{|I(u)|+n}
\label{eq:R_K}
\end{dmath}

\begin{dmath}
P(B_j=b_j | R_u=k)=\frac{|I(B_j=b_j,R_u=k)|+1}
{|I(R_u=k)|+m}
\label{eq:R_bj}
\end{dmath}

\noindent where $|I(u)|$ is the number of items that the user $u$ rated, $|I(R_u=k)|$ is the number of items that the user $u$ gives the rating equal to $k$, and $|I(B_j=b_j,R_u=k)|$ is the number of ratings $k$ that the user $u$ gave to items with the attribute of $b_j$. We add one in the numerator and $n$ as the range of ratings and $m$ as the range of attribute value in the denominators, because of the Laplace estimate that helps us in avoiding strong probabilities. 

Subsequently, we need to find the item acceptance probability which is $P(R_i=k|A=a_u)$. It implies the general acceptance of item $i$ from users like user $u$. For example, if two reviewers are similar to each other and one of them rated ``The Godfather'' 5, we want to know how likely is that the other one gives the same rating. Again, by naive Bayesian assumption, we have equation~\ref{eq:item_pref}.

\begin{dmath}
P\left (R_i=k | A=a_i \right)
= \frac{ P(R_i=k) \times P(A_1,A_2, \dots ,A_n | R_i=k)}
{P(A_1,A_2, \dots ,A_n)} \\
= \frac{ P(R_i=k) \times \prod_{j=1}^{j=m}  P(A_j | R_i=k)}
{P(A_1,A_2, \dots ,A_n)} , A_j \in \{A_1,A_2, \dots ,A_m \}
\label{eq:item_pref}
\end{dmath}

\noindent where $P(R_i=k)$ is the prior probability that item $i$ receives a rating value $k$, and $P(A_j | R_i=k)$ is the conditional probability that a user has attribute $A_j$ equal to $a_j$ given that she rates item $i$ as $k$. Note that in the previous equations, both $P(B_1,B_2, \dots ,B_n)$ and $P(A_1,A_2, \dots ,A_m)$ are normalizing constants. 

Finally, the influence from immediate friends should be obtained, i.e. $P(R_{u,i}=k | \{  R_{v,i}=r_{v,i}: \forall v \in U_i \cap N_u \})$. Some methods use the correlation between the user and its neighbors based on user attributes, but this correlation is hard to capture with a simple similarity or correlation function. Then the authors in\cite{he2010social} suggest that we can use the histogram of the differences between the immediate friends rating and the user rating. Therefore, for each user $u$ and her neighbor $v$, we have equation~\ref{eq:hist}:
\begin{equation}
P(R_{u,i}=k|R_{v,i}=r_{v,i}) \propto H(k-r_{v,i})
\label{eq:hist}
\end{equation}

In order to calculate it for all the neighbors of $u$, these differences are multiplied and divided by a normalization factor of the  histogram of each immediate friend pair. 

\section{Link Prediction for Social Networks}
Online social media is growing with a significantly important pace. 
An applicative domain of social-based recommender system is link prediction on social media~\cite{al2011survey}. Want et al.~\cite{DBLP:journals/corr/WangXWZ14} provide a thorough review over this topic. They divide the link recommendation on social networks to two major categories: similarity-based approach and learning-based approach. They also explore the social theory-based metrics, the node-based metrics and the topology-based metrics. The latter ones mainly considers the neighbors and the path for qualification metrics.

A number of papers has focused on particular social networks. For example, Yao et al.~\cite{yao2011context} explore the friend suggestion in online photo-sharing communities such as Facebook and flicker. In another article, Liben-Nowell et al.~\cite{epasto2015ego} explores the problem in the context of freind suggestion over Twitter. As a future work, emojis can be considered as a tuner for link prediction techniques, because the emoji usage analysis shows regularities~\cite{fede2017representing,seyednezhad2017understanding} and semantics~\cite{seyed2018flairs} on Twitter users. Thus, the user with similar feelings and common friend may be subjected for link suggestions. 

If we consider the networks of researchers as a social network (e.g. Mendeley, ResearchGate, etc.), then research paper recommendation may be treated as a social-based recommendation system.
Because one of the major applications of the recommender systems is to recommend a set of relevant and useful papers to a scholar in the right time.
In addition to the time limitations, the issue of copyright prevent a recommender system to access to the full content of a paper. 
Two popular approaches are context-based collaborative filtering~\cite{liu2015context} and co-citation~\cite{mcnee2002recommending}. In the former, the authors use the network of citations to create the rating matrix. The latter takes into account the assumption that if two papers cite the same papers, they are similar. 

Haruna et al.~\cite{haruna2017collaborative} propose a collaborative method that uses the public data about the paper for the recommendation purposes. In their method, if author {\it A} writes a paper {\it P}, they consider recommending papers that have two conditions: They are co-cited with the paper {\it P} of the author {\it A}, and have common references with paper {\it P}. They show their method outperforms context-based collaborative filtering and co-citation techniques. Another method for research paper recommendation is to analyze the topics of the papers~\cite{pan2010research}.

\chapter*{ACKNOWLEDGMENTS}
\label{sec:Acknowledgments}
The authors would like to thank the NSF for funding the AMALTHEA REU and Florida Institute of Technology for hosting the program. The authors would also like to acknowledge support from the NSF grant No. \href{http://www.nsf.gov/awardsearch/showAward?AWD_ID=1560345}{1560345}. Any opinions, findings, and conclusions or recommendations expressed in this material are those of the authors and do not necessarily reflect the views of the NSF.
\addcontentsline{toc}{chapter}{Acknowledgments}

\addcontentsline{toc}{chapter}{Bibliography}

\bibliographystyle{abbrv}

\end{document}